\documentstyle{spacekap}
\begin{opening}
\title{TEV GAMMA-RAYS FROM PROTON BLAZARS}
\author{Karl Mannheim}
\institute{Universit\"ats-Sternwarte, Geismarlandstra{\ss}e 11,
D-37083 G\"ottingen, Germany\newline
Internet: kmannhe@uni-sw.gwdg.de}
\end{opening}
\begin{document}
\begin{abstract}
Proton acceleration in nearby blazars
can be diagnosed measuring their intense TeV $\gamma$-ray emission.
Flux predictions for 1101+384 (Mrk421) and 1219+285 (ON231), both strong
EGRET sources ($0.1-10$~GeV), are obtained from model
spectra of unsaturated
synchrotron pair cascades fitted to publicly available multifrequency
data.  An experimental effort to confirm
the predicted emission in the range 1-10 TeV would be of great
importance for the problems of the origin of cosmic rays, the era of
galaxy formation and the cosmological distance scale.
\end{abstract}
\section{Introduction}
\subsection{Relativistic jets}
It came as a big surprise to most of the community when numerous
powerful extragalactic $\gamma$-ray sources were discovered with
the Energetic Gamma Ray Experiment Telescope EGRET onboard the Compton
Gamma-Ray Observatory CGRO (Fichtel et al. 1994, von Montigny
et al. 1994).
The
sources were soon identified as blazars, a subclass of active
galactic nuclei (AGN), which are kwown to be the most powerful and rapidly
variable nonthermal emitters also in other energy bands
(Bregman 1990). The most intriguing fact
about the discovery was that the received energy flux from many of these
enigmatic objects is dominated by $\gamma$-rays.
The precursor experiment COS-B
had already indicated the existence of such $\gamma$-ray dominated
extragalactic sources, baptized `proton quasars' by Moffat et al. (1983).
The TeV detection of Mrk421 by the Whipple group makes it probable
that the blazar $\gamma$-ray spectra extend to still higher energies.
%
%
%Even more bewildering than the relative luminosities were the
%absolute apparent luminosities.
%Mrk421 at a distance of $120(H_\circ/{\rm
%75 km/s/Mpc})^{-1}$ Mpc is the
%only object other than the Crab nebula at a distance of $5$ kpc
%detected by the Whipple Cher\'enkov telescope.  This implies that
%the TeV $\gamma$-ray power of Mrk421 is more than $10^8$ times that
%of the Crab - tacitly assuming that the distance to Mrk421 can be inferred
%%from from its redshift via Hubble expansion of
%the universe.
Although it is already difficult to conceive that
$\gamma$-ray sources as powerful as
Mrk421 with $L_\gamma\simeq 10^{43}$ erg
s$^{-1}$ exist in nature, some blazars
detected by EGRET had apparent $\gamma$-ray luminosities
of even
$L_\gamma\approx 10^{48}$ erg s$^{-1}$.

These new challenging observations can be explained assuming that
the $\gamma$-rays are emitted in a lighthouse beam directed towards
the observer (McBreen 1979).
The beaming pattern, and thus the blazar phenomenon,
arises from the relativistic bulk motion of plasma
jets ejected from the vicinity of a supermassive black hole
(Blandford \& K\"onigl 1979).   Relativistic radio
jets are commonly observed in flat-spectrum radio sources and superluminal
expansion of the radio structure indicates bulk motion with
Lorentz factors $\gamma_{\rm j}=10-100$
(Begelman et al. 1994).
Whereas most particles in the jet plasma remain in
thermal equilibrium,
a small fraction of them is accelerated to high energies
retaining isotropy in the comoving frame by
pitch-angle scattering.
Most probably, particles gain energy by repeatedly
crossing collisionless shock fronts in the supersonic jet flow
(Drury 1983, Begelman \& Kirk 1990).
A power law energy
distribution indicates the stochastic nature of the acceleration mechanism.
\subsection{The radiation process responsible for $\gamma$-ray emission}
The high particle energies and
the twenty orders of magnitude in frequency of the emitted
nonthermal radiation are reminiscent of the properties
of cosmic rays in our own
Galaxy where electrons produce synchrotron radio emission and
protons with Lorentz factors up to $10^{11}$ generate energetic $\gamma$-rays
by pion production
$p+p\rightarrow \pi^\circ+X$ and their subsequent decay.
Proton-matter interactions
can not be the radiation mechanism
responsible for the variable and highly luminous
$\gamma$-ray emission from blazars,
since this would require unrealistically large amounts of target
matter in the jets.  It would also imply a significant enhancement
of light-elements due to spallation of heavier
nuclei which has not been observed (Baldwin et al. 1979).
However, it was noticed
by Blumenthal (1970) and Colgate (1983)
that the compact
radiation fields in extragalactic radio sources represent
by far the most important
target for relativistic protons
(cf. also references in Berezinsky et al. 1990).
At very high energies protons scatter inelastically
by photo-production of secondaries, i.e.
$p+\gamma\rightarrow \pi^\circ+p$ and $p+\gamma\rightarrow e^++e^-+p$.
In contrast
to the low-density interstellar medium where $\gamma$-rays can escape
freely, blazars absorb the energetic $\gamma$-rays
resulting from the decay of secondaries
via pair creation $\gamma+\gamma\rightarrow
e^++e^-$.  As a consequence,
unsaturated pair cascades
develope which attenuate the $\gamma$-ray flux above the
energy $\epsilon^*$ where $\tau_{\gamma\gamma}(\epsilon^*)=1$,
but which let $\gamma$-rays with energies $\epsilon<\epsilon^*$
escape freely (Mannheim et al. 1991).

There has been some debate whether inverse-Compton scattering
or synchrotron radiation is the dominant energy loss process
of pairs in the blazar jets.  In radio jets the observed
energy density
ratio of  photons and  magnetic field, which
governs the ratio of the cooling time scales,
is  $a=u_\gamma/u_B<1$ at a scale of $r=1$~kpc
(Harris et al. 1994).  Assuming a tangled advected magnetic field,
which scales as $B\propto r^{-1}$,
and noting that the photon energy density scales
as $u_\gamma\propto r^{-2}$ for a constant
comoving frame luminosity  it follows that $a<1$ remains invariant.
This is in accord with X-ray measurements
of radio sources (Biermann et al. 1988).

Theoretically $a<1$ is expected, if the target photons are produced  locally in
the jet plasma.   The particles with energy density $u_{\rm rel}$
produce the radiation with energy density
$u_\gamma\le u_{\rm rel}$.
Presumably, particles are accelerated by scattering
off magnetic field irregularities with energy density
$u_{\delta B}\le u_B$, hence $u_\gamma\le u_{\rm rel}<u_B$.
If, on the other hand, the relativistic
particle pressure would exceed the magnetic field pressure,
the particles would break out of their magnetic confinement
immediately turning-off further acceleration.
%
%Observations of radio through optical synchrotron spectra require
%typical maximum
%Lorentz factors of the radiating electrons
%$\gamma_{\rm e}\approx 10^{4}B^{-1/2}$ where
%$B$ denotes the magnetic field strength in cgs units  (Bregman 1990).

Only if the particles responsible for the $\gamma$-ray emission
are accelerated very close to the AGN,
so that the energy density
of the external radiation field exceeds that instrinsic to the jet,
inverse-Compton scattering becomes the dominant cooling mechanism.
Sikora et al. (1994) have argued that the re-radiation of a central thermal
UV spectrum by the broad emission line clouds can produce such
a strong external radiation field, if the $\gamma$-ray production
occurs within the central parsec of the AGN
powering the $\gamma$-ray blazar.  Dermer et al. (1992) have argued that direct
thermal radiation from  an accretion disk
represents the majority of target photons.
It was predicted by the latter model that
spectral breaks greater than $\Delta s=0.5$
should not occur in the $\gamma$-ray regime
-- in contrast to the observations.
Also required is the presence of a source
of thermal target radiation dominating the bolometric luminosity,
although no indication of this big blue bump spectral
component is seen in the spectra of several blazars with a
near-infrared/optical
cut-off
(Bregman 1990). It is a fundamental question, whether or
not such a thermal emission component is ubiquitous
in AGN.  Indeed, the emission component can remain undiscovered
due to  Doppler boosting of the nonthermal luminosity
by factors $>10^4$
or due to obscuration by dust.
If, however, obscuration is responsible for hiding the thermal source,
it would also be hidden for the relativistic particles in the jet.
Consequently,  nonthermal optical outbursts should not occur
simultanously with $\gamma$-ray
outbursts.
Another limitation of these models
is that the extreme inverse-Compton losses postulated imply that
an unknown impulsive electron acceleration mechanism must be operating.
Magnetic reconnection could be such a process.  However,
the thickness of neutral sheets is probably insufficient
to obtain the large potential drops required for the
great values of the electron Lorentz factors.
It is also unclear, whether  the
observed power law electron spectra can be produced in this way.
\subsection{Proton-induced cascades}
In the most simple model for blazar emission it is assumed
that steady, conical relativistic jets of Lorentz factor $\gamma_{\rm j}$
with opening angle $\Phi\approx \gamma_{\rm j}^{-1}$
are viewed at a small angle $\theta$
(Blandford \& K\"onigl 1979, Mannheim 1993a).
Photons emitted isotropically in the comoving
frame are blueshifted with the Doppler factor
$\delta=[\gamma(1-\beta\cos\theta)]^{-1}$
and the apparent luminosity exceeds the
comoving frame luminosity by a large margin.
The outflow is assumed to advect a tangled magnetic field
and relativistic particles in equipartition with the
magnetic field.  Most probably shock acceleration
compensates adiabatic losses during the expansion.
Synchrotron-self-absorption frequency, break frequeny
(above which the cooling time scale is shorter than the dynamical
time scale) and emissivity obey a simple  scaling
with distance $r$ along the jet, such that the superposition
of spectra from scales between $r_{\rm min}$ and $r_{\rm max}$ yields
a flat spectrum $S_\nu\propto const.$ up to the break frequency
$\nu_{\rm b}\approx 10^{11}-10^{13}$~Hz above which the spectrum
steepens by one power yielding $S_\nu\propto \nu^{-1}$.
In the near-infrared/soft X-ray regime the primary electron
synchrotron spectrum turns over steeply when the cooling
time scale becomes shorter than the acceleration time scale.
For protons, the break energy is reached at
\begin{equation}
\gamma_{\rm p,b}=\left(m_{\rm p}\over m_{\rm e}\right)^3
{1+a\over 1+240a}\gamma_{\rm e,b}
\approx 3\cdot 10^9 \left(\gamma_{\rm e,b}\over 100\right)
\end{equation}
Note that $a=u_\gamma/u_B$ refers only to the synchrotron
photon density in the radio through X-ray range, since higher
energy photons are unimportant as a target, for
inverse-Compton scattering because of the Klein-Nishina
cross-section and for photo-production because of the threshold
energy.
The proton break energy is much greater than  the electron break energy,
since the proton energy losses are negligible compared
to the electron energy losses at the same energy.
The proton-induced luminosity for
$\gamma_{\rm p}\le\gamma_{\rm p,b}$
has the value
\begin{equation}
L_{\rm casc}\simeq \eta \left(
{\gamma_{\rm p}\over \gamma_{\rm p,b}}\right)L_{\rm syn}
\end{equation}
where $\eta=u_{\rm p}/u_{\rm e}$ denotes the energy density ratio of
relativistic protons and electrons, respectively.
It is very likely that protons do not always reach
the extremely high energy $\gamma_{\rm p}=
\gamma_{\rm p,b}$.
One limitation arises from the condition that the proton Larmor
radius must remain smaller than the shock radius which is itself
less than the jet radius (Bell 1978).
Hence it follows that for a universal cosmic ray proton/electron ratio
$\eta$ it is the proton maximum energy which determines the
$\gamma$-ray spectrum.

The brightest $\gamma$-ray emission
is expected from the most compact parts of the jet where most of
the nonthermal target photons with $\nu\ge \nu_{\rm b}$
are produced, i.e. at
the jet length $r=r_{\rm b}$ (jet radius
$r_{\rm b,\perp}\approx r_{\rm b}\Phi$). At smaller
scales $r<r_{\rm b}$ synchrotron-self-absorption and rapid
dynamical evolution darken
the jet, whereas at larger scales $r>r_{\rm b}$
the radio flux increases and
the infrared photon flux decreases.
Due to the threshold condition for secondary particle production, e.g.
$\nu_{\rm th}\ge 2\cdot 10^{12} (\gamma_{\rm p}/10^{10})^{-1}$~Hz for
pions, only the infrared and higher frequency photons are
relevant as a target for the protons.
Indeed, the jet would
be $\gamma$-ray bright also on large scales
(steep spectrum radio sources), if the proton
maximum energy were allowed to  reach up to a value $\gamma_{\rm p}
\ge \gamma_{\rm p,b}$ at any scale in the jet.
However, this is impossible due to
the Larmor radius constraint and the finite acceleration time scale.
The greatest values of the proton
maximum energy seem to be achieved in the hot spots of radio galaxies
where $\gamma_{\rm p,max}\approx 3\cdot 10^{11}\approx .07 \gamma_{\rm p,b}$
(Biermann \& Strittmatter 1987).
Notice, however,
that blazars with $L_\gamma\approx 100 L_{\rm syn}$
require $\gamma_{\rm p}=\gamma_{\rm p,b}\approx 10^{10}$ ($\eta=100$)
in the compact parts of the blazar jet.  Some of these
protons escape after isospin flip $p+\gamma\rightarrow n+\pi^+$
as neutrons .  The neutrons then suffer $\beta$-decay at a distance
of $d_{\rm n}=\tau_{\rm n}c\gamma_{\rm n}\approx 100$~kpc from the blazar.
In this way an extragalactic proton flux without
adiabatic losses is injected into the intergalactic medium.
In the observer's frame these particles have a Lorentz factor
$\gamma\approx \delta \gamma_{\rm p,b}\approx 10^{11}$.

Tacitly
it is assumed that the jets become radiative at a rather
{\em large} distance from their origin.  Shocks are generated
as the jet propagates through the external
steep pressure gradient of the elliptical host galaxy.
This could happen at a distance
as far as $1$~kpc from the kinematical center (Sanders 1983).

The spectrum of the proton-induced cascades is
complex, producing broken power laws and double-humped
spectra between X-ray energies and 100~TeV
(Mannheim et al. 1991).
The internal absorption of $\gamma$-rays in  'proton
blazars' (Mannheim 1993a), which determines the
precise shape of the cascade spectrum, is difficult
to assess, since the photon density is highly inhomogenous,
e.g. varying from the limb to the center of the jet and
varying with length along the jet.
However, the {\it effective} compactness is strongly constrained once
the cascade spectrum is known
from observations between keV and GeV photon energies.

Since proton-induced cascades inject electromagnetic energy
at ultra-high energies,
it is, in fact, the strongest implication of the proton blazar
model that TeV $\gamma$-ray emission can be regarded
as a typical
property.  This is in marked contrast to the leptonic
models where physical parameters have to be pushed
to their limits.
The best candidate sources for TeV detection
are $\gamma$-ray sources
with a hard X-ray spectral index $\alpha_{\rm x}=0.5-0.7$ indicating
a low
comoving frame
radiation compactness.
\subsection{Merits of $\gamma$-ray measurements}
External absorption of $\gamma$-rays by the
infrared/optical radiation field produced inside the blazar
host galaxy (Mannheim 1993b)
and by the cosmic background radiation produced by all galaxies
(Gould \& Schr\'eder 1966, Stecker et al. 1992, MacMinn \& Primack 1994)
must be taken into account.  This limits
the number of possible TeV sources
to the few nearest blazars, but it opens up
new and challenging possibilities for probing the era
of galaxy formation and the cosmic distance scale, since
$\tau_{\gamma\gamma}(\epsilon_\gamma)\propto
n_{\rm ir}(\epsilon_\gamma) H_\circ^{-1}$ where
$n_{\rm ir}(\epsilon_\gamma)$ is the near-infrared
photon density at the pair creation threshold
$\epsilon_\circ=2(m_{\rm e}c^2)^2/\epsilon_\gamma\simeq
0.5/(\epsilon_\gamma/ {\rm TeV})~{\rm eV}$.
A prerequisite for measuring either $H_\circ$ or $n_{\rm ir}$
via the exponential cut-off produced by
external absorption is a theoretical prediction of the unabsorbed $\gamma$-ray
spectra.  To this end I discuss below the $\gamma$-ray spectra of 1219+285 and
Mrk421
in some detail.  The strategy is to find combined electron synchrotron
spectra and cascade spectra fitting to the multifrequency data for the
same set of parameters.  In this way the spectrum is highly constrained
yielding firm predictions for the TeV range.
\section{TeV emission from Mrk421 and 1219+285}
Internal absorption of $\gamma$-rays inside the blazar jets
is crucial to the shape of the emerging radiation spectrum.
For a steady jet the optical depth in the comoving frame
is given by
\begin{equation}
\tau_{\gamma\gamma}=\int_{2(m_{\rm e}c^2)^2/\epsilon_\gamma}^\infty
n_{\rm syn}(\epsilon)\sigma_{\gamma\gamma}(\epsilon,\epsilon_\gamma)
r_{\rm b,\perp}d\epsilon\simeq {aB^2\epsilon_\gamma\over 8\pi
\ln[\epsilon_{\rm c}/\epsilon_{\rm b}](m_{\rm e}c^2)^2}
{\sigma_{\rm T}\over 3}r_{\rm b,\perp}
\end{equation}
where
$a\simeq (1+\eta)^{-1}\beta_{\rm j}\gamma_{\rm j}\Phi\approx
(1+\eta)^{-1}$.
Above the comoving frame energy $\epsilon^*$ where $\tau_{\gamma\gamma}(
\epsilon^*)=1$ the cascade spectrum steepens by one power,
since
\begin{equation}
I_\gamma=I_\circ {1-\exp{[-\tau_{\gamma\gamma}]}\over \tau_{\gamma\gamma}}
\rightarrow I_\circ \epsilon_\gamma^{-1}\ \ (\tau_{\gamma\gamma}\gg 1)
\end{equation}
{}From the proton blazar model fits shown in Figs.~1,2 obtained for the
parameters
in Tab.~1 one can infer that the turnover energies
in the observer's frame
$\delta\epsilon_\gamma^*$ are of the order of TeV, viz.
\begin{equation}
\epsilon_\gamma^*({\rm obs})\approx
\left(\delta\over 10\right)
\left(\ln[\epsilon_{\rm c}/\epsilon_{\rm b}]\over 6\right)
\left(\eta\over 100\right)\left(B\over 10~{\rm G}\right)^{-2}
\left(r_{\rm b,\perp}\over 3\cdot 10^{15}~{\rm cm}\right)^{-1}
{}~{\rm TeV}
\end{equation}
It is the length scale $r=r_{\rm b}$
where the jet becomes optically thin for infrared radiation which
regulates the internal absorption to values such that TeV is the
typical turnover energy.  Another typical energy can be found
as a direct
consequence:  Pairs produced from the optically thick part of the
cascade at $\epsilon_\gamma^*$ have Lorentz factors $\gamma_{\rm e}^*=
\epsilon_\gamma^*/2$.  The synchrotron photons from these pairs
have characteristic energies
\begin{equation}
\epsilon_\gamma^{**}({\rm obs})=3\cdot 10^{-14}\delta B\left(\gamma_{\rm e}^*
\right)^2\approx 0.02 \left(\delta\over 10\right)^{-1}\left(B\over 10~{\rm G}
\right)\left(\epsilon_\gamma^*({\rm obs})\over {\rm TeV}
\right)^2~{\rm MeV}
\end{equation}
Below this energy the unsaturated cascade is optically thin,
so that the typical X-ray spectral index is $\alpha_{\rm X}=0.5-0.7$
in marked contrast to saturated cascades with $\alpha_{\rm X}=0.9-1.0$
(Svensson 1987).

No definite predictions for the proton blazar spectra
during flux outbursts can be made at present, since this
would require to solve non-stationary
cascade equations and to couple the radiative transport
with a dynamical model for shock propagation and particle
acceleration.  Most probably, outbursts represent
shocks propagating into the optically thin
part of the jet.
The optical outbursts
(primary electrons) and the hard X-/$\gamma$-ray
outbursts (cascade pairs)
should be correlated, since the cascades develope
in the synchrotron photon target produced by the primary
electrons.
The
high-energy cascade emission therefore follows
the optical outburst
on the very short
proton photo-production cooling time scale.
This time scale is larger than the optical
outburst decay time scale by a factor of $\approx \eta/R$ where
$R=L_\gamma/L_\circ$ denotes the $\gamma$/optical luminosity ratio, cf. Eq.(2).
If there is a universal value of
the proton/electron ratio $\eta$,
it is predicted that the sources with weakest $\gamma$-ray
emission compared to the optical emission display the
largest delays in $\gamma$-rays.
\begin{table}[h]
\caption{Physical parameters for the proton blazar
model fits shown in Figs.~1,2.  See Mannheim (1993a)
for further explanations}
\begin{center}
\begin{tabular}{lll}\hline
	&  1219+285 &  Mrk421 \\
\hline\\
B [G]   &     4     &   40    \\
$\eta$  &     30    &   100    \\
$r_{\rm b\perp}$ [cm] & $7\cdot 10^{15}$&$2\cdot 10^{15}$\\
$\gamma_{\rm p}$ & $2\cdot 10^9$& $6\cdot 10^7$\\
$\delta$ &  $7$   &   $31$  \\
$\gamma_{\rm j}$&  $5$  &  $20$  \\
$\theta$ [deg] &  $7$  & $1.5$  \\
$L_{44}$  &  $30$  & $18$  \\
$\gamma_{\rm p}/\gamma_{\rm p,b}$&  $3\cdot 10^{-2}$& $5\cdot 10^{-4}$\\
\hline\end{tabular}
\end{center}
\end{table}  \begin{figure}
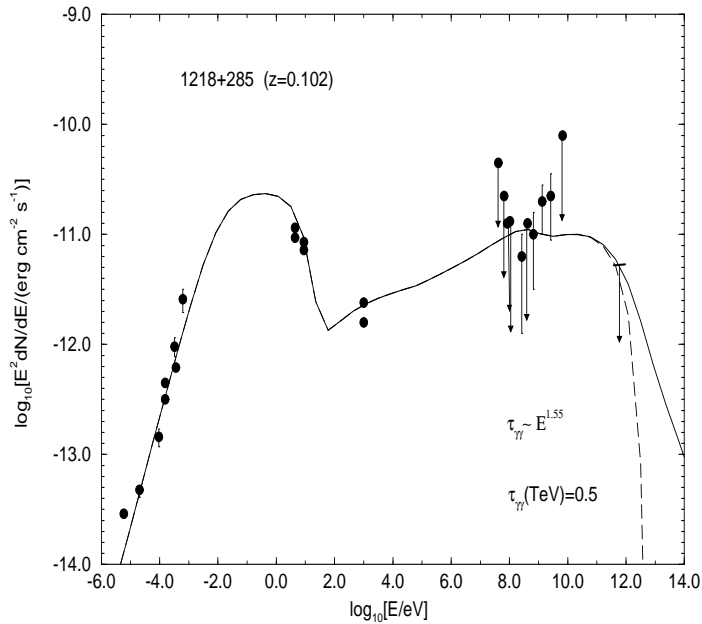

\vspace{10cm}
%\centerline{\psfig{figure=1219.ps,height=10cm,width=8.8cm}}
\caption{Multifrequency spectrum of 1219+285.
Data were obtained from the NED, von Montigny et al. (1994),
Fink (priv.com.)
and the Whipple group (priv.com.).  The solid line
shows the proton blazar model fit for the parameter values in
Tab.~1.  The dashed line shows the expected flux accounting for
external absorption (Stecker et al. 1992)}
\end{figure}
\begin{figure}
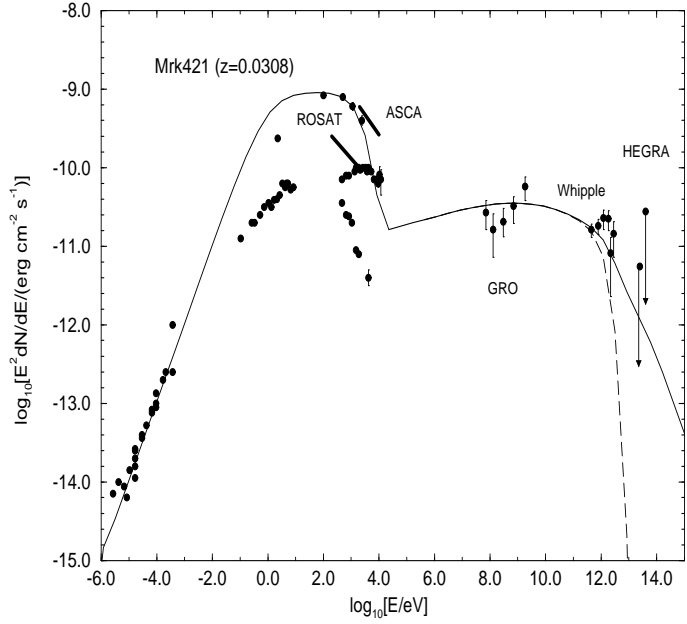

\vspace{10cm}
%\centerline{\psfig{figure=mrk421.ps,height=10cm,width=8.8cm}}
\caption{Multifrequency spectrum of Mrk421
obtained from the NED, von Montigny
et al. (1994), Punch et al. (1992), Thomas (priv.com.),
K\"uhn (1994) and Karle (1994).  The solid line
shows the proton blazar model fit for the parameter
values in Tab.~1.  The dashed line shows the expected flux accounting for
external absorption (Stecker et al. 1992) }
\end{figure}

\section{Discussion and conclusions}
Abundaning geocentrism it is hard to conveive that
baryonic cosmic rays should {\it not} be present
in radio jets.
Requiring the Larmor radii of protons to be smaller than the
jet radius yields a
maximum energy of up to $10^{11}$~GeV -- large enough
to make extragalactic radio jets the source of cosmic rays
above EeV (Rachen \& Biermann 1993).

Recent $\gamma$-ray measurements in the energy band
$0.1-10$~GeV have established that the
brightest and most compact extragalactic jets
are indeed powerful cosmic particle accelerators with
their
maximum luminosity temporarily in the $\gamma$-ray band.
Proton-induced unsaturated synchrotron cascades can
be responsible for the observed $\gamma$-ray emission.
In marked contrast to the relativistic protons, for
which energy losses become important only
near their maximum energy, electrons suffer
severe energy losses during the entire acceleration process
which hinders them to reach the high Lorentz factors
needed to produce ample TeV photons
by inverse-Compton scattering of UV-photons.
Therefore, TeV $\gamma$-ray emission can be considered
as a typical indicator of proton acceleration in radio jets.

However, nature has limited our view of the universe
by a very restricted horizon
for $\gamma$-ray eyes.
Due to pair-absorption of $\gamma$-rays on
cosmic background radiation fields
the TeV world ends at a luminosity distance of roughly
$400$~Mpc.   Only the few nearest blazars are therefore
potential TeV sources.  Nevertheless, the precise value
of the TeV horizon depends on two quantities of
superior astrophysical interest:  The Hubble-constant
and the near-infrared photon density which entails
the conditions at birth of normal galaxies.
We may hope to shed more light on these cosmological
cornerstones
by measuring and analysing the TeV $\gamma$-ray spectra of nearby
blazars.

Neglecting the non-simultanity of the
multifrequency data and
on the basic assumptions of the proton blazar model
(biconical
relativistic outflow,
isotropy of ultra-relativistic particles in the
comoving frame,
equipartition of relativistic particle energy and
magnetic field energy) I was able
to obtain flux predictions
in the air Cher\'enkov energy range
and above for two nearby blazars
Mrk421 ($z=0.308$) and 1218+285 ($z=0.102$).
%Both objects were strong EGRET sources and Mrk421
%has been observed by the Whipple telescope.
%1219+285 had a very hard spectrum in the EGRET band
%($s=1.4\pm0.4$), but has not yet been observed
%above $0.3$~TeV.

In the TeV range
internal absorption of $\gamma$-rays producing
a break in the power-law spectrum competes with
external absorption producing an exponential cut-off.
The observed TeV $\gamma$-ray emission from
Mrk421 is consistent with little external
absorption, whereas significant external absorption is likely
to be present in the spectrum of 1219+285.
If the Hubble-constant had a value
$H_\circ\ge 75$~km/s/Mpc or if the near-infrared
photon density had a value less than the expected
value $n_\gamma(0.5{\rm eV})\approx 2\cdot 10^{-3}$~cm$^{-3}$,
the proton-induced cascade flux would have been above
the Whipple upper limit for this source.
However, since external absorption does not
decrease the observed flux in the air Cher\'enkov energy range by a large
factor,
1219+285 is very likely to be the next
blazar detected by a Cher\'enkov telescope.
\begin{acknowledgements}
S. Westerhoff is acknowledged for helpful assistance.
This research has made use of the NASA/IPAC Extragalactic Database
(NED).
\end{acknowledgements}
{}

\begin{thebibliography}{}
\bibitem[]{}  Baldwin, J., Boksenberg, A., Burbidge, G.,
Carswell, R., Cowsik, R., Perry, J., Wolfe, A., 1977, AA, {\bf 61}, 165
\bibitem[]{}  Begelman, M.C., Kirk, J.G., 1990, ApJ, {\bf 353}, 66
\bibitem[]{}  Begelman, M.C., Rees, M.J., Sikora, M., 1994, ApJ, {\bf 429}, L57
\bibitem[]{}  Bell, A.R., 1978, MNRAS, {\bf 182}, 443
\bibitem[]{} Berezinsky, V.S., Bulanov, S.V., Dogiel, V.A.,
Ginzburg, V.L. (ed.), Ptuskin,  V.S., Astrophysics of Cosmic Rays, North
Holland, Amsterdam, 1990
\bibitem[]{}  Biermann, P.L., Strittmatter, P.A., 1987, ApJ., {\bf 322}, 643
\bibitem[]{}  Biermann, P.L., et al., 1988, AA, {\bf 185}, 9
\bibitem[]{}  Blandford, R.D., K\"onigl, A., 1979, ApJ, {\bf 232}, 34
\bibitem[]{}  Blumenthal, G.R., 1970, Phys.ReV. D1, No.6, 1596
\bibitem[]{}  Bregman, J.N., 1990, AAR, 125
\bibitem[]{} Colgate, S.A., 1983, 18th Int. Cosmic Ray Conf.,  Vol.2,
p. 230
\bibitem[]{}  Dermer, C.D., Schlickeiser, R., Mastichiadis, A., 1992,
AA, {\bf 256}, L27
\bibitem[]{}  Drury, L O'C, 1983, Rep.Prog.Phys., {\bf 46}, 973
\bibitem[]{}  Fichtel, C.E., et al., 1994, ApJ, in press
\bibitem[]{}  Gould, R.J., Schr\'eder, G., 1966, Phys.Rev.Lett., {\bf 16}, 252
\bibitem[]{}  Harris, D.E., Carilli, C.L., Perley, R.A., 1994,  Nat,
{\bf 367}, 713
\bibitem[]{}  Karle, A., and HEGRA collaboration, 1994,
PhD-thesis, University of M\"unchen
\bibitem[]{}  K\"uhn, M., 1994, PhD-thesis, University of Kiel
\bibitem[]{}  Mannheim, K., Kr\"ulls, W., Biermann, P.L., 1991,
AA, {\bf 251}, 723
\bibitem[]{}  Mannheim, K., 1993a, AA, {\bf 269}, 67
\bibitem[]{}  Mannheim, K., 1993b, Phys.Rev.D, {\bf 48}, 5270
\bibitem[]{}  MacMinn, D., Primack, J., 1994, ApJ, submitted
\bibitem[]{}  McBreen, B., 1979, AA, {\bf 71}, L17
\bibitem[]{}  Moffat, A.F.J., Schlickeiser, R., Shara, M.M., Sieber, W.,
Tuffs, R., K\"uhr, H., 1983, ApJ, {\bf 271}, L45
\bibitem[]{}  von Montigny, C., et al., 1994, ApJ, accepted for publication
\bibitem[]{}  Punch, C.W., et al., 1992, Nat, {\bf 358}, 477
\bibitem[]{}  Rachen, J.P., Biermann, P.L., 1993, AA, {\bf 272}, 161
\bibitem[]{}  Sanders, R.H., 1983, ApJ, {\bf 266}, 73
\bibitem[]{}  Sikora, M., Begelman, M.C., Rees, M.J., 1994, ApJ, {\bf 421}, 153
\bibitem[]{}  Schlickeiser, R., 1984, ApJ, {\bf 277}, 485
\bibitem[]{}  Stecker, F.W., de Jager, O., Salamon, M.H., 1992, ApJ, {\bf 390},
L49
\bibitem[]{}  Svensson, R., 1987, MNRAS, {\bf 227}, 403
\end{thebibliography}
\end{document}